\documentclass[conference]{IEEEtran}

\usepackage{graphicx}
\usepackage{comment}
\usepackage{xcolor}

\ifCLASSINFOpdf
\else
\fi
\begin{document}

\title{Creating, Using and Assessing a Generative-AI-Based Human-Chatbot-Dialogue Dataset with User-Interaction Learning Capabilities}

\author{
\IEEEauthorblockN{Alfredo Cuzzocrea\textsuperscript{\textsection}}
\IEEEauthorblockA{iDEA Lab\\ University of Calabria\\
Rende, Italy \\
\& Department of Computer Science\\
University of Paris City\\
Paris, France\\
alfredo.cuzzocrea@unical.it}
\and
\IEEEauthorblockN{Giovanni Pilato}
\IEEEauthorblockA{ICAR-CNR\\
National Research Council of Italy\\
Palermo, Italy\\
giovanni.pilato@icar.cnr.it}
\and
\IEEEauthorblockN{Pablo G. Bringas}
\IEEEauthorblockA{Faculty of Engineering\\ University of Deusto\\
Bilbao, Spain\\
pablo.garcia.bringas@deusto.es}}

\maketitle
\begingroup\renewcommand\thefootnote{\textsection}
\footnotetext{This research has been made in the context of the Excellence Chair in Big Data Management and Analytics at University of Paris City, Paris, France.}
\endgroup

\begin{abstract}
The study illustrates a first step towards an ongoing work aimed at developing a dataset of dialogues potentially useful for customer service conversation management between humans and AI chatbots. 
The approach exploits ChatGPT 3.5 to generate dialogues. One of the requirements is that the dialogue is characterized by a specific language proficiency level of the user; the other one is that the user expresses a specific emotion during the interaction. The generated dialogues were then evaluated for overall quality. The complexity of the language used by both humans and AI agents, has been evaluated by using standard complexity measurements. Furthermore, the attitudes and interaction patterns exhibited by the chatbot at each turn have been stored for further detection of common conversation patterns in specific emotional contexts.
The methodology could improve human-AI dialogue effectiveness and serve as a basis for systems that can learn from user interactions.
\end{abstract}

\IEEEpeerreviewmaketitle

\section{Introduction}

Emotions play a key role during Human-Robot and Human-Computer Interaction (HRI/HCI), and applying Affective Computing (AC) methods to identify and assess human emotional states makes machines potentially more acceptable to human users \cite{pilato2018data,cuzzocrea2021composite,moin2023emotion}.
The difficulties in computational approaches for emotion detection arise from the fact that emotions are complex, and interpreting them is often a function of the context in which they are exhibited. 

In recent years, the NLP community has been particularly interested in Emotion Recognition in Conversation (ERC). 
On the other hand, awareness of emotions is also crucial for dialogue generation and management, because it involves knowledge of the emotional state of the interlocutor, and this knowledge can be leveraged to increase the level of empathy and better manage the dialogue \cite{poria2019emotion}.

There are numerous contributions to the literature on the development of datasets for emotion recognition.
Most of them focus on Ekman's primary emotions \cite{ekman1994all}.
Some examples are 
CrowdFlower's Emotion Dataset \cite{Crowdflower2016}, Friends \cite{joshi-etal-2016-harnessing} , EmoBank \cite{buechel-hahn-2017-emobank}. Moreover, considerable methodologies have been developed to construct datasets through the utilization of publicly available sources, such as newspapers, online dialogues, and social media platforms. This approach enables researchers to realize repositories of language data, facilitating the creation of emotion recognition datasets, e.g. SemEval-2018 Task 1: Affect in Tweets (AIT-2018) \cite{SemEval2018Task1}, 
Sentiment140 \cite{goel2016real}, Emotion Intensity Dataset (EmoInt) \cite{mohammad-bravo-marquez-2017-wassa}, the International Survey on Emotion Antecedents and Reactions (ISEAR) \cite{scherer1994evidence}. Other approaches exploit movies, e.g. the Stanford Sentiment and Emotion Classification (SSEC) \cite{schuff2017annotation,mohammad2017stance} or physiological signals, e.g. The Database for Emotion Analysis using Physiological Signals (DEAP) \cite{koelstra2011deap}. 
Popular datasets are also the Interactive EMOtional dyadic motion CAPture database (IEMOCAP), collected by the Speech Analysis and Interpretation Laboratory (SAIL) at the University of Southern California (USC) \cite{busso2008iemocap}, and DailyDialog, a dataset of dialogues about different topic of routine conversations,  characterized by human-written and plain language; in this dataset, data were associated with communication intention and emotion information \cite{li2017dailydialog}.

On the other hand, Large Language Models (LLMs) have also been used to emulate conversations. As an example, Li et al. \cite{li-etal-2016-deep}  present a model that simulates conversations between two virtual agents, employing policy gradient methods to reward sequences that exhibit desirable conversational properties. The authors of \cite{steindl2023generating} investigated the use of  a prompt-following Large Language Model (LLM) to augment existing datasets for training Task-Oriented Dialogue (TOD) systems. Exploiting the strengths of both human-generated data and machine-generated data, potentially leading to more robust and accurate models for real-world dialogue applications.Chen et al. \cite{chen2023places} presented PLACES, a comprehensive framework designed to generate synthetic dialogues by including topic-specific information,  background context, and expert-written conversations as training examples.
DIALOGIC \cite{li2022controllable} introduced a novel methodology for controllable dialogue simulation, which enables the generation of synthetic dialogues that are annotated with Dialog State Tracking (DST) labels. 
The Language Models as Data, LAD, as proposed by Mehri et al. \cite{mehri2022lad}, leverages a novel paradigm to produce synthetic dialogues that are characterized by structural richness and variability. 
SynthDST \cite{kulkarni2024synthdst} represents a domain-agnostic framework for producing synthetic dialogue datasets that are enriched with Dialog State Tracking (DST) annotations.

Recognizing emotions becomes essential in customer care handling. In most cases, it becomes crucial to design systems that swiftly recognize the user's emotion and the language level employed by the client to produce an effective automatic interaction or to design artificial systems that can assist a human agent in effectively and adequately interacting with the client.

We have chosen to use ChatGPT 3.5 to produce some examples of interactions between people in specific circumstances, expressing certain emotions that we want to emphasize and utilizing a determined level of linguistic complexity. In particular, we selected the following emotions: \textit{joy}, \textit{sadness}, \textit{anger}, \textit{fear}, \textit{surprise}, \textit{disgust}. Furthermore, we decided to require that the user was using one of the following \textit{Common European Framework of Reference for Languages} (CEFR) levels during the dialogue: \textit{A2}, \textit{B2}, and \textit{C2}.

Along with these directions, ChatGPT entirely generated each dialogue. Furthermore, ChatGPT associated a label denoting the emotional attitude of the Agent or the user for each turn of the generated conversation. This makes it possible to extract and save the emotional dynamics involved during each interaction: utilizing the dialogue structure generated by ChatGPT, it is feasible to develop a system that parses and annotates these interactions according to predefined criteria, thereby facilitating the creation of realistic exemplars of human-computer interaction over time. 

The generated dialogues were analyzed using specific measures of linguistic complexity, thereby enabling a comprehensive evaluation of their efficacy and relevance. The conversations were subsequently archived in a repository for future querying purposes.  

The remainder of this paper is organized as follows: Section II illustrates the proposed approach, Section III provides the experimental results, Section IV reports conclusions and future works.

\section{The Proposed Approach}
The illustrated work uses ChatGPT to generate a collection of short conversations that can be tailored to a specific scenario. Each dialogue is characterized by a different level of text complexity and a specific predominant emotion, reflected by the user's tone during the interaction, to match real human interactions.

The generated dialogues were preliminary evaluated before being included into the dataset to assess their accuracy and adherence to the predetermined contextual parameters, the desired linguistic complexity, and the emotional content. 
This process was aimed at discarding those dialogues that failed to meet the specifications,  ensuring the quality and consistency of the resulting dataset. 

The study leveraged the capabilities of OpenAI's ChatGPT-3.5 Generative AI to conduct experiments within a controlled context. Specifically, we simulated a scenario in which a client talks with a customer service representative from a hypothetical phone company, while expressing one of the six fundamental emotions identified by Eckman\cite{ekman1999basic}. For each generated conversation, we asked ChatGPT to produce a both a set of dialogues where the user explicitly and implicitly conveyed a specific emotion. In the latter case, we exploited appropriate prompts to prevent the LLM to use words directly expressing the target emotion. We named the conversations arising from this procedure as \textit{Implicit Emotion Dialogues} (IED) that were also systematically labeled and analyzed, analogously to the procedure illustrated in \cite{asta2023supporting}.

The procedure associates two labels to each generated dialogue in the dataset: an emotional label and a CEFR language level.
The prompt provided to ChatGPT specified the creation of short, interactive dialogues consisting of approximately five turns between a customer and an agent in a customer service context.

A thorough evaluation of the generated dialogues was conducted through an interactive process, ensuring that the language employed was suitable and the overall quality of interaction met expectations. The annotated dialogues can be utilized as a knowledge base to train an interactive system capable of acquiring effective interaction habits. To achieve this goal, explicit emotional cues and reactions provided as metadata within the generated dialogues were leveraged to identify recurring patterns and sequences for further analysis.

To generate a specific language complexity level, we referred in the prompt to the Common European Framework of Reference for Languages (CEFR). It is a paradigmatic framework for categorizing language proficiency, comprising six discrete levels of linguistic ability (\textit{A1} to \textit{C2}). These levels are stratified into three categories: \textit{Basic User} (\textit{A} category), \textit{Independent User} (\textit{B} category), and \textit{Proficient User} (\textit{C} category).
The CEFR's schema is characterized by the employment of \textit{can-do} descriptors, which provide explicit definitions of what a learner can accomplish at each level. 
The schema offers a detailed analysis of the cognitive and linguistic skills involved in language acquisition. The framework's descriptions are universal, applicable to all languages, and they provide a clear trajectory for progression through each skill set, with the six levels (A1, A2, B1, B2, C1, C2) reflecting an increasing mastery of linguistic competence. 

The generated dialogues were analyzed to be sufficiently sure of including in the dataset only dialogues that adhere to the required characteristics. In particular, we have considered the coherence between the required emotion expressed by the user and the request to ChatGPT, naming it \textit{emotional coherence}. Its value is \textit{true} or \textit{false}

Furthermore, we subsequently examined each generated dialogue to assess the congruence between the expected linguistic complexity used by the user and the generated dialogue. We named this parameter: 'language complexity coherence'. The value assumed by this quantity is boolean (i.e., \textit{true}/\textit{false}), and it exploited a set of language complexity measures that were performed in the dialogue as a double check for the generated conversations.

Finally, we considered a qualitative measurement of the general \textit{Quality of Interaction} (QoI) inherent to the generated dialogues, assigning one of three distinct scores to each dialogue: \textit{Sufficient} (S), \textit{Adequate} (A), or \textit{Fail} (F). A QoI value of \textit{S} indicates that the language exhibited is natural and effectual in conveying the intended emotional resonance and linguistic complexity. Conversely, an \textit{A}-score implies that while the language may lack complete naturalness, it remains acceptable and effectively conforms to its purpose, even if it may contain words that do not accurately represent the intended emotion or align with the requested linguistic complexity. In contrast, a QoI value of \textit{F} indicates that the generated dialogue is characterized by confusion or unusual language that fails to accurately reflect the intended emotional resonance and the requested linguistic complexity. Any dialogue to which a QoI value of \textit{F} was assigned was evaluated as unsuitable for analysis and excluded from further consideration and storage in the repository.

Readability assessment entails determining the ease with which a text can be understood and processed. This process typically involves attributing an appropriate reading level to the text, which can be helpful for diverse levels of readers and second-language learners alike. Various measures have been proposed in the literature to assess a text's readability or, conversely, difficulty level.

To facilitate analysis of the generated dialogues, we leveraged the Automatic Readability Tool for English (ARTE) \cite{ARTE}, which enables the automatic computation of multiple readability metrics for texts. The selection of ARTE was due to its free and user-friendly access to the computation of different metrics on input texts.

Specifically, the generated dialogues were subjected to readability measurements to evaluate their understandability and verify that ChatGPT had maintained the specified language complexity.

For the sake of completeness, we now list the measures employed in our analysis and available in the ARTE Tool. These metrics were computed to determine the readability of user-generated and agent-generated content within the dialogues and assess the overall difficulty of the language used.

\begin{itemize}
    \item \textit{Flesch Reading Ease}: it employs a quantitative approach to assess the complexity of written texts, assigning higher scores to passages that are more challenging to comprehend and lower scores to those that are more straightforward. The evaluation framework takes into account several linguistic parameters, including: the total number of syllables present in the text, the number of words employed in the passage, the number of sentences composing the text.
    \item \textit{Flesch-Kincaid Grade Level}: it is a widely recognized and established readability formula that quantitatively estimates the reading proficiency required to comprehend a given text. This assessment is based on two primary factors: the average sentence length, which evaluates the syntactic complexity of the written passage, and the complexity of the words employed. The resultant scores provide a standardized measure corresponding to U.S. grade levels. Initially developed for educational applications, this metric has been increasingly applied in various contexts.
    \item \textit{Automated Reading Index}: The Automated Readability Index (ARI) is another well-known readability test for English texts \cite{kincaid1975derivation}. The ARI score leverages the characters-per-word metric as an alternative to the syllable-per-word approach. This paradigm shift enables a more practical computational process, since character counting is generally more efficient and accurate than syllable computation.
    \item \textit{New Dale-Chall Readability Formula}: it is a quantitative measure of text readability designed to evaluate the level of comprehension difficulty presented by a given passage. The revised version of this formula employs a comprehensive lexicon including approximately 3,000 words that are typically within the comprehension of average fourth-grade American students.
The words not included in this list are considered cognitively demanding and potentially challenging for readers to understand \cite{chall1995readability}. 
    \item \textit{CAREC - Crowdsourced Algorithm of Reading Comprehension}: it takes into account different parameters, like the average age of acquisition (Kuperman) for all content words, the bigram range score (COCA) for all words, the average trigram proportion score (BNC-written) for all words, the average imageability score (MRC) for all content words, the average frequency score (Brown) for all words, the average type token ratio of lemma trigrams for all trigrams, the proportion of lemma types that occur in the next paragraph for all paragraphs, the number of temporal connectives divided by number of words in text, the proportion of noun lemma types that occur in the next paragraph for all paragraphs, the number of content word lemma types, the positive adjective scores derived from four different corpora, the average standard deviation of word length for all words, and the average character entropy for all characters \cite{ARTE}.        
    \item \textit{CAREC\_M - Crowdsourced Algorithm of Reading Comprehension Modified}: The CAREC\_M variant is the same than the CAREC model, except that the factor Number of content word lemma types divided by number of content words is considered instead of Number of content word lemma types \cite{ARTE}.
    \item \textit{Coh-Metrix L2 Readability Index (Approximated)}: this model takes into account the number of sentences in text, the average frequency score (SUBTLEXus) for all content words logged, the proportion of noun and pronoun lemma types that occur in the next two sentences for all sentences \cite{ARTE};
    \item \textit{Sentence-BERT}: it is a deep-learning-based readability model which is based on SBERT \cite{reimers2019sentence}\cite{ARTE}.    
\end{itemize}

\begin{figure*}
    \centering
    \includegraphics[scale=0.21]{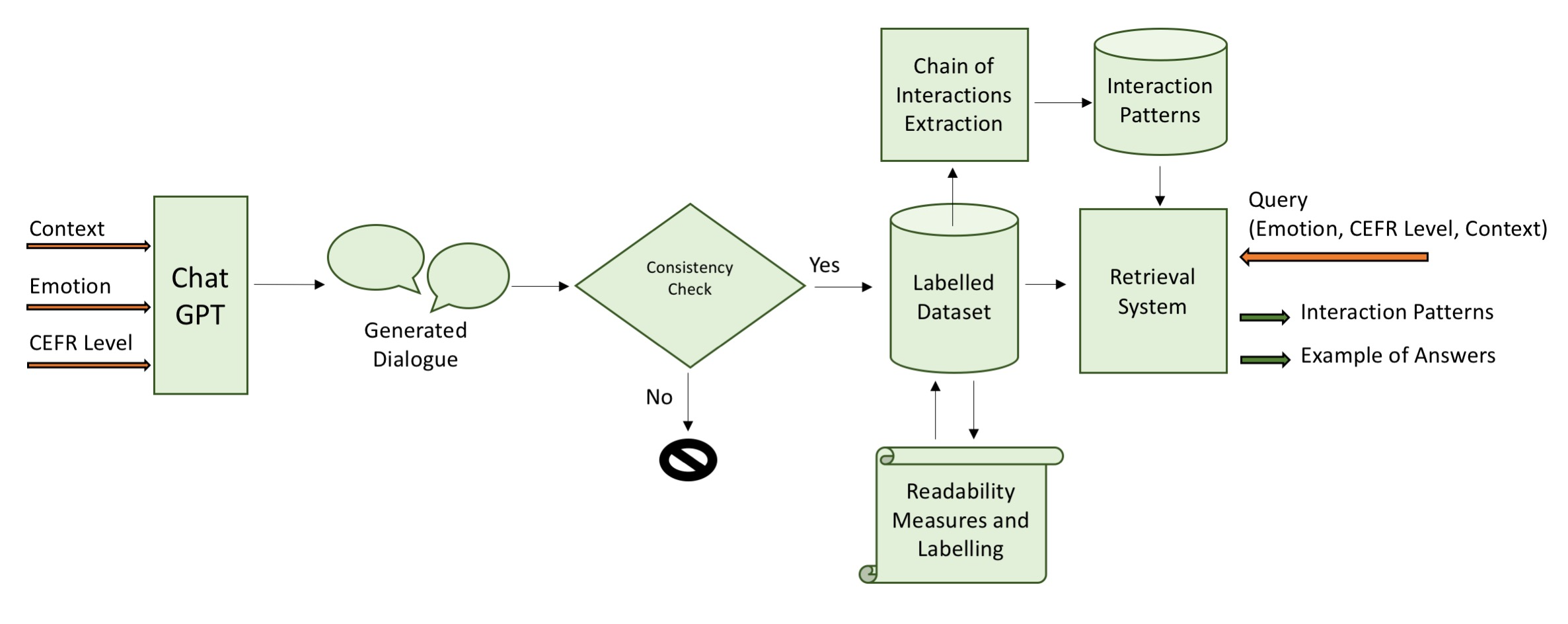}
    \caption{The overall schema of the proposed approach}
    \label{fig:ARI}
\end{figure*}

\section{Experimental Results}
A set of experiments were conducted by using ChatGPT 3.5. In particular, we asked the bot to produce some conversational scenarios, each one designed to evoke a specific emotional answer from the user. 

Below, we report some generated dialogues, categorized by two features: the emotion and the linguistic complexity expressed by the user.

In the conversations where specific phone brands or models were referenced, we have anonymized the references by replacing them with the string \textit{Brand Model}, to avoid any potential misrepresentation of actual commercial products.

To facilitate future analysis, the dialogue patterns and their related chains of attitude labels saved in a repository. Each sequence 
was linked to relevant information, like the emotional attitude of the user and his CEFR language level. This makes it possible an organization and comparison of typical interaction patterns for future interactions by artificial agents.

In what follows, we report some examples regarding the \textit{anger} and \textit{surprise} emotions. For each emotion, we show three dialogues with different CEFR levels, namely \textit{A2}, \textit{B2}, and \textit{C2}. Furthermore, we illustrate the readability results obtained with different literature measures to automatically, even if roughly, test the coherence with the language proficiency required in the prompts.

\subsection{Anger and A2 CEFR Language Level}
In the following dialogue, the client expresses anger using language appropriate for the A2 CEFR level. The agent responds calmly and offers assistance to resolve the issue. Between parenthesis it is shown the emotion expressed by the subject that has been assigned by ChatGPT.

\noindent \textbf{Client} (\textit{angry}): Hello, I'm calling about my phone, and I'm not happy at all!\\
\textbf{Agent} (\textit{calm}): I'm sorry to hear that. Can you tell me what seems to be the problem?\\
\textbf{Client} (\textit{frustrated}): Yeah, my phone is not working, and I need it for work. It's really annoying!\\
\textbf{Agent} (\textit{understanding}): I understand. Let's try to figure this out. Can you tell me what specifically is wrong with your phone?\\
\textbf{Client} (\textit{irritated}): It won't turn on! I've tried everything, and it's just dead.\\
\textbf{Agent} (\textit{reassuring}): I'm sorry you're experiencing this issue. Can you confirm the make and model of your phone, please?\\
\textbf{Client} (\textit{agitated}): It's a \textit{Brand Model}.\\
\textbf{Agent} (\textit{confirming}): Thank you. Let's troubleshoot this together. Have you tried charging your phone to see if it responds?\\
\textbf{Client} (\textit{frustrated}): Yes, I've tried that a hundred times! Nothing works!\\
\textbf{Agent} (\textit{sympathetic}): I understand your frustration. Let's schedule a technician to assist you. Can I have your preferred time for a visit?\\
\textbf{Client} (\textit{grudgingly}): Fine. Anytime after 5 p.m. works for me.\\
\textbf{Agent} (\textit{apologetic}): I apologize for the inconvenience. We'll have someone there after 5 p.m. Thank you for your patience.\\

From the above interaction, it is moreover inferred the following chain of interaction: [\textit{(Client , angry) $\rightarrow$ (Agent , calm) $\rightarrow$ (Client , frustrated) $\rightarrow$ (Agent , understanding) $\rightarrow$ (Client , irritated) $\rightarrow$ (Agent , reassuring) $\rightarrow$ (Client , agitated) $\rightarrow$ (Agent , confirming) $\rightarrow$ (Client , frustrated) $\rightarrow$ (Agent , sympathetic) $\rightarrow$ (Client , grudgingly) $\rightarrow$ (Agent , apologetic) }]

\subsection{Anger and B2 CEFR Language Level}
In this dialogue, the client expresses anger using language appropriate for the B2 CEFR level. The agent responds with attentiveness and offers assistance to address the issue.

\noindent \textbf{Client} (\textit{angry}): Good afternoon. I'm calling because I'm quite frustrated with my phone situation.\\
\textbf{Agent} (\textit{attentive}): I'm sorry to hear that. Can you please explain what's going on with your phone?\\
\textbf{Client} (\textit{irritated}): Well, my phone refuses to turn on, and I need it urgently for work. It's incredibly inconvenient.\\
\textbf{Agent} (\textit{understanding}): I understand the urgency. Let's work together to resolve this. Can you provide details on what you've tried so far to fix the issue?\\
\textbf{Client} (\textit{exasperated}): I've attempted multiple troubleshooting methods, but nothing seems to work. It's just completely unresponsive.\\
\textbf{Agent} (\textit{sympathetic}): I'm sorry for the frustration this has caused. To assist you further, can you confirm the make and model of your phone?\\
\textbf{Client} (\textit{aggravated}): It's a \textit{Brand Model}.\\
\textbf{Agent} (\textit{confirming}): Thank you for providing that information. Let's explore additional troubleshooting steps. Have you tried charging your phone to see if it responds?\\
\textbf{Client} (\textit{frustrated}): Yes, I've charged it, but there's still no sign of life. I'm at my wit's end.\\
\textbf{Agent} (\textit{empathetic}): I completely understand your frustration. Let's expedite the resolution process. Would you be available for a technician to visit today?\\
\textbf{Client} (\textit{reluctantly}): I suppose so. Anytime after 5 p.m. would work for me.\\
\textbf{Agent} (\textit{apologetic}): I apologize for the inconvenience. We'll arrange for a technician to assist you after 5 p.m. Thank you for your patience.

From the above interaction, it is moreover inferred the following chain of interaction: 
[\textit{(Client , angry) $\rightarrow$ 
(Agent , attentive) $\rightarrow$ 
(Client , (irritated) $\rightarrow$ 
(Agent , understanding) $\rightarrow$ 
(Client , exasperated) $\rightarrow$ 
(Agent , sympathetic) $\rightarrow$ 
(Client , aggravated) $\rightarrow$ 
(Agent , confirming) $\rightarrow$ 
(Client , frustrated) $\rightarrow$ 
(Agent , empathetic) $\rightarrow$ 
(Client , reluctantly) $\rightarrow$ 
(Agent , apologetic)}]

\subsection{Anger and C2 CEFR Language Level}
In the following dialogue, the client expresses anger using language appropriate for the C2 CEFR level. The agent responds attentively and offers assistance to address the issue.

\noindent \textbf{Client} (\textit{angry}): Good afternoon. I'm contacting you because I'm deeply frustrated by the ongoing issue with my phone.\\
\textbf{Agent} (\textit{attentive}): I'm sincerely sorry to hear that. Could you please elaborate on the nature of the problem?\\
\textbf{Client} (\textit{irritated}): Certainly. My phone has become unresponsive, which is highly inconvenient, particularly for my work obligations.\\
\textbf{Agent} (\textit{understanding}): I understand the urgency. Let's work together to resolve this promptly. Could you detail the troubleshooting steps you've already taken?\\
\textbf{Client} (\textit{exasperated}): I've exhausted various troubleshooting methods to no avail. Despite multiple attempts, my phone remains entirely non-functional.\\
\textbf{Agent} (\textit{sympathetic}): I'm truly sorry for the inconvenience this has caused. To proceed, could you confirm the make and model of your phone?\\
\textbf{Client} (\textit{aggravated}): It's a \textit{Brand Model}.\\
\textbf{Agent} (\textit{confirming}): Thank you for providing that information. Let's explore further options. Have you attempted to charge your phone to determine if it responds?\\
\textbf{Client} (\textit{frustrated}): Yes, I've attempted charging it, yet there's been no indication of life. This situation is becoming increasingly infuriating.\\
\textbf{Agent} (\textit{empathetic}): I completely empathize with your frustration. Let's expedite the resolution process. Are you available for a technician to visit today?\\
\textbf{Client} (\textit{reluctantly}): Regrettably, yes. Anytime after 5 p.m. would be suitable for me.\\
\textbf{Agent} (\textit{apologetic}): I apologize for the inconvenience. We'll arrange for a technician to assist you after 5 p.m. Thank you for your understanding.

From the above interaction, it is moreover inferred the following chain of interaction: 
[\textit{(Client , angry) $\rightarrow$ 
(Agent , attentive) $\rightarrow$ 
(Client , irritated) $\rightarrow$ 
(Agent , understanding) $\rightarrow$ 
(Client , exasperated) $\rightarrow$ 
(Agent , sympathetic) $\rightarrow$ 
(Client , aggravated) $\rightarrow$ 
(Agent , confirming) $\rightarrow$ 
(Client , frustrated) $\rightarrow$ 
(Agent , empathetic) $\rightarrow$ 
(Client , reluctantly) $\rightarrow$ 
(Agent , apologetic)}]

\subsection{Surprise and A2 CEFR Language Level}
In this dialogue, the client expresses surprise using language appropriate for the A2 CEFR level. The agent responds with curiosity and offers assistance to address the unexpected situation.

\noindent \textbf{Client} (\textit{surprised}): Um, hi. I'm calling because I'm, like, really surprised right now. My phone's acting strange.\\
\textbf{Agent} (\textit{curious}): Oh, I see. Can you tell me what's happening with your phone?\\
\textbf{Client} (\textit{amazed}): Yeah, it's, like, not turning on! I didn't expect this at all.\\
\textbf{Agent} (\textit{concerned}): That does sound surprising. Let's see what we can do. Can you confirm the make and model of your phone, please?\\
\textbf{Client} (\textit{hesitantly}): Um, sure. It's a \textit{Brand Model}.\\
\textbf{Agent} (\textit{confirming}): Got it. Let's try a few troubleshooting steps. Have you tried charging your phone to see if it responds?\\
\textbf{Client} (\textit{surprised}): Oh, um, no, I haven't. I'll try that now.\\
\textbf{Agent} (\textit{encouraging}): Great. Let's see if that helps. Please let me know if anything changes.\\
\textbf{Client} (\textit{surprised}): Wow, it's turning on now! I didn't think that would work.\\
\textbf{Agent} (\textit{reassuring}): I'm glad to hear that. Sometimes, it's the simple things that do the trick. If you encounter any more issues, don't hesitate to reach out.

From the above interaction, it is moreover inferred the following chain of interaction:[\textit{(Client , surprised) $\rightarrow$ 
(Agent , curious) $\rightarrow$ 
(Client , amazed) $\rightarrow$ 
(Agent , concerned) $\rightarrow$ 
(Client , hesitantly) $\rightarrow$ 
(Agent , confirming) $\rightarrow$ 
(Client , surprised) $\rightarrow$ 
(Agent , encouraging) $\rightarrow$ 
(Client , surprised) $\rightarrow$ 
(Agent , reassuring)}]

\subsection{Surprise and B2 CEFR Language Level}
In this dialogue, the client expresses surprise using language appropriate for the B2 CEFR level. The agent responds with curiosity and offers assistance to address the unexpected situation.

\noindent \textbf{Client} (\textit{surprised}): Hello. I'm calling because I'm quite surprised by the current situation with my phone.\\
\textbf{Agent} (\textit{curious}): I understand. Can you please explain what's happening?\\
\textbf{Client} (\textit{amazed}): Well, my phone suddenly won't turn on. It's really unexpected.\\
\textbf{Agent} (\textit{concerned}): That does sound surprising. Let's see if we can resolve it. Could you confirm the make and model of your phone?\\
\textbf{Client} (\textit{hesitant}): Of course. It's a \textit{Brand Model}.\\
\textbf{Agent} (\textit{confirming}): Thank you. Let's try a few troubleshooting steps. Have you attempted to charge your phone to see if it responds?\\
\textbf{Client} (\textit{surprised}): Oh, actually, I haven't tried that yet. Let me give it a go.\\
\textbf{Agent} (\textit{encouraging}): Alright, let's see if that helps. Please keep me updated on any changes.\\
\textbf{Client} (\textit{surprised}): Wow, it's turning on now! I didn't expect that to work.\\
\textbf{Agent} (\textit{reassuring}): I'm glad to hear that. Sometimes, simple solutions can be quite effective. If you encounter any further issues, feel free to contact us again.

From the above interaction, it is moreover inferred the following chain of interaction:
[\textit{(Client , surprised) $\rightarrow$ 
(Agent , curious) $\rightarrow$ 
(Client , amazed) $\rightarrow$ 
(Agent , concerned) $\rightarrow$ 
(Client , hesitant) $\rightarrow$ 
(Agent , confirming) $\rightarrow$ 
(Client , surprised) $\rightarrow$ 
(Agent , encouraging) $\rightarrow$ 
(Client , surprised) $\rightarrow$ 
(Agent , reassuring)}]

\subsection{Surprise and C2 CEFR Language Level}
In this dialogue, the client expresses surprise using language appropriate for the C2 CEFR level. The agent responds with attentiveness and offers assistance to address the unexpected situation.

\noindent \textbf{Client} (\textit{surprised}): Good day. I'm contacting you because I find myself quite taken aback by the current state of my phone.\\
\textbf{Agent} (\textit{inquiring}): I understand. Could you please elaborate on what exactly is happening?\\
\textbf{Client} (\textit{amazed}): Certainly. My phone has suddenly become unresponsive, which is truly unexpected.\\
\textbf{Agent} (\textit{concerned}): That does sound surprising. Let's work to resolve it. Could you confirm the make and model of your phone?\\
\textbf{Client} (\textit{hesitant}): Certainly. It's a \textit{Brand Model}.\\
\textbf{Agent} (\textit{confirming}): Thank you. Let's proceed with troubleshooting. Have you attempted to charge your phone to ascertain if it responds?\\
\textbf{Client} (\textit{surprised}): Actually, I haven't considered that. I'll give it a try now.\\
\textbf{Agent} (\textit{encouraging}): Very well, let's see if that yields any results. Please keep me informed of any developments.\\
\textbf{Client} (\textit{surprised}): Remarkably, it's powering on now! I hadn't anticipated such a simple solution.\\
\textbf{Agent} (\textit{reassuring}): I'm glad to hear that. Sometimes, the most straightforward approaches prove to be the most effective. If any further issues arise, do not hesitate to contact us.

From the above interaction, it is moreover inferred the following chain of interaction:
[\textit{(Client , surprised) $\rightarrow$ 
(Agent , inquiring) $\rightarrow$ 
(Client , amazed) $\rightarrow$ 
(Agent , concerned) $\rightarrow$ 
(Client , hesitant) $\rightarrow$ 
(Agent , confirming) $\rightarrow$ 
(Client , surprised) $\rightarrow$ 
(Agent , encouraging) $\rightarrow$ 
(Client , surprised) $\rightarrow$ 
(Agent , reassuring)}]

\subsection{Readability Results}

All dialogues created with ChatGPT and that overcame the consistency check, i.e. that were judged emotionally coherent and with a sufficient or at least adequate quality of interaction, were aggregated by the CEFR level required in the prompt, regardless of the emotions the user was asked to manifest. 
After setting the CEFR level, we considered separately the sentences attributed to the user and the sentences attributed to the agent.
Readability was then measured by CEFR level and by the role played in the dialogue.

Individual turns of the considered interlocutor were then randomly extracted. Each turn was appended to a list of sentences until the total number of words did not reach the maximum number allowed by the ARTE online tool (specifically, 1000 words). 
Of course, the last turn cannot be interrupted. As a consequence, if the number of words in the last turn to be included in the analysis is such that it exceeds the maximum allowed number of words, the specific is not included in the list of sentences to be evaluated.
In the end, a text of approximately 950 words is built and evaluated according with the different readability measures provided by the ARTE tool.
This operation was carried out several times for each CEFR level and each interlocutor type (User or Agent). Since the sentences were taken randomly, the text considered for the analysis is varied, and it includes different emotions that ChatGPT was asked to express by the user. This because we are focused on analyzing the text readability independently of the specific emotions expressed in the conversation.

\begin{figure}
    \centering
    \includegraphics[scale=0.21]{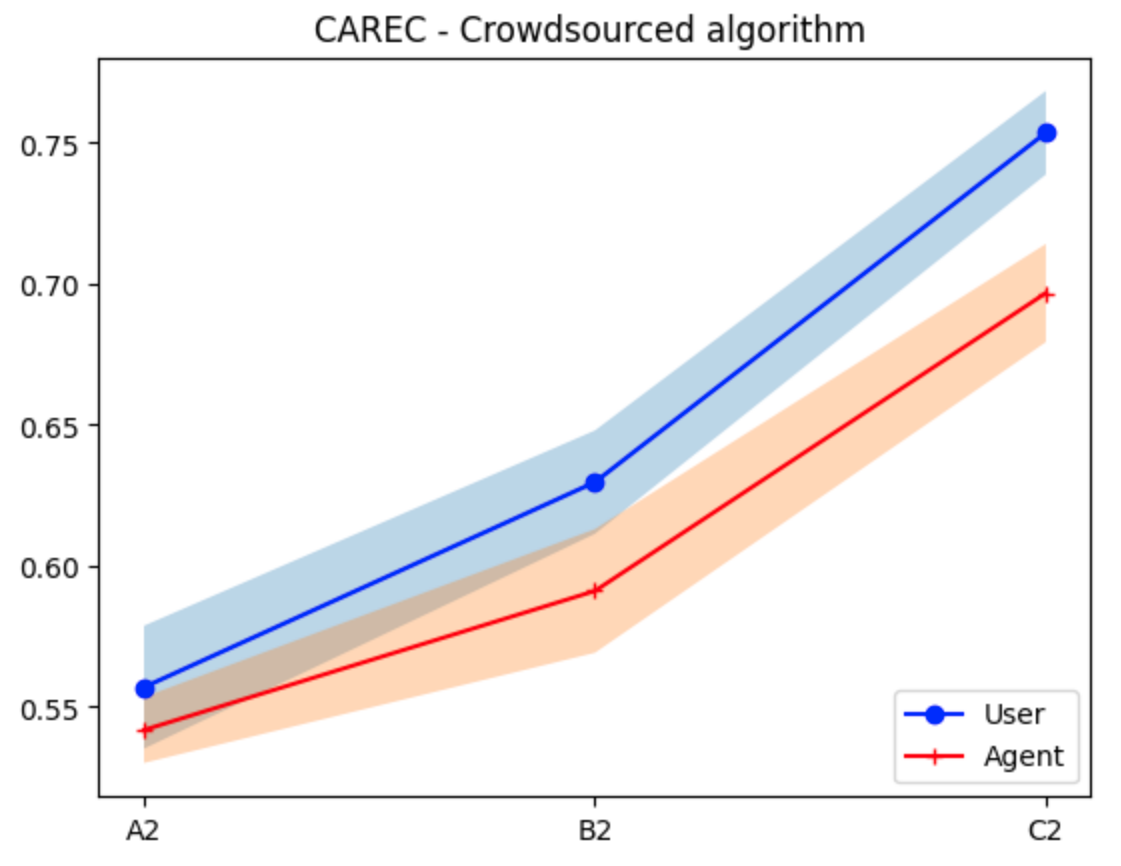}
    \caption{The CAREC average readability results for the A2, B2, and C2 CEFR levels  both for the User and the Agent}
    \label{fig:CAREC}
\end{figure}

\begin{figure}
    \centering
    \includegraphics[scale=0.21]{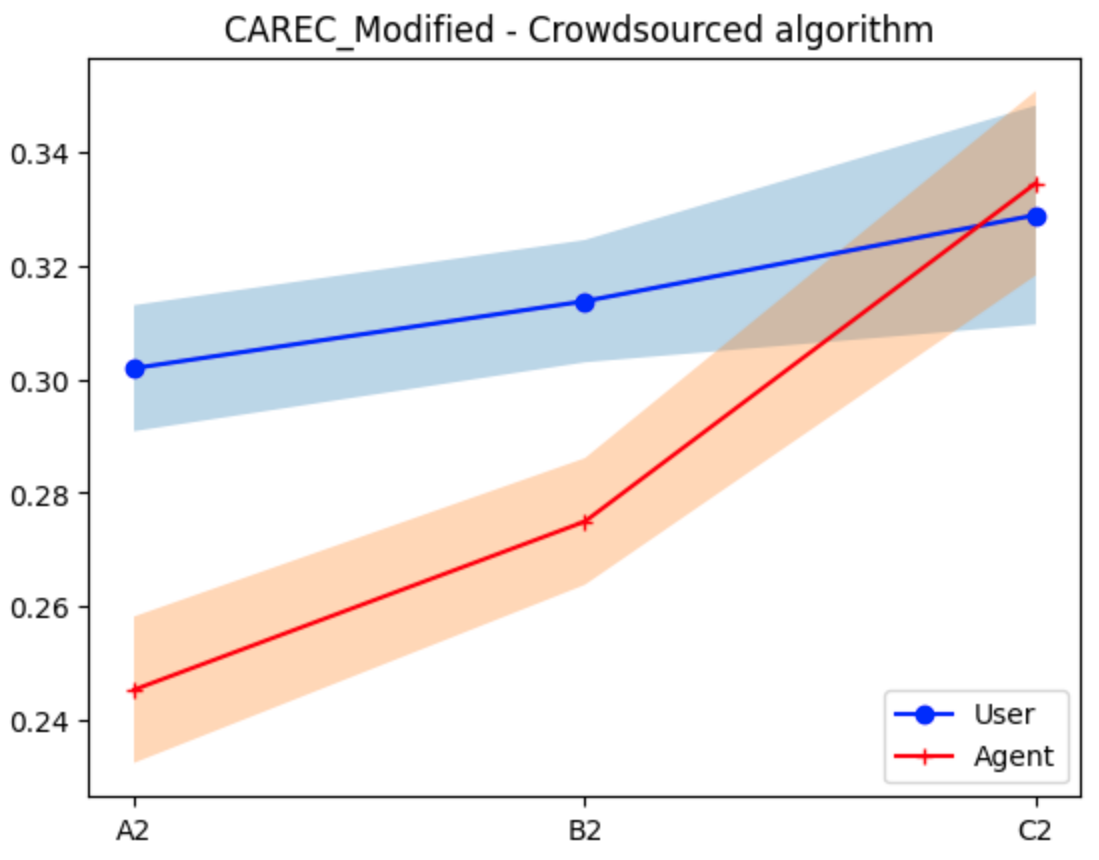}
    \caption{The CARECM average readability results for the A2, B2, and C2 CEFR levels  both for the User and the Agent}
    \label{fig:CARECM}
\end{figure}

\begin{figure}
    \centering
    \includegraphics[scale=0.21]{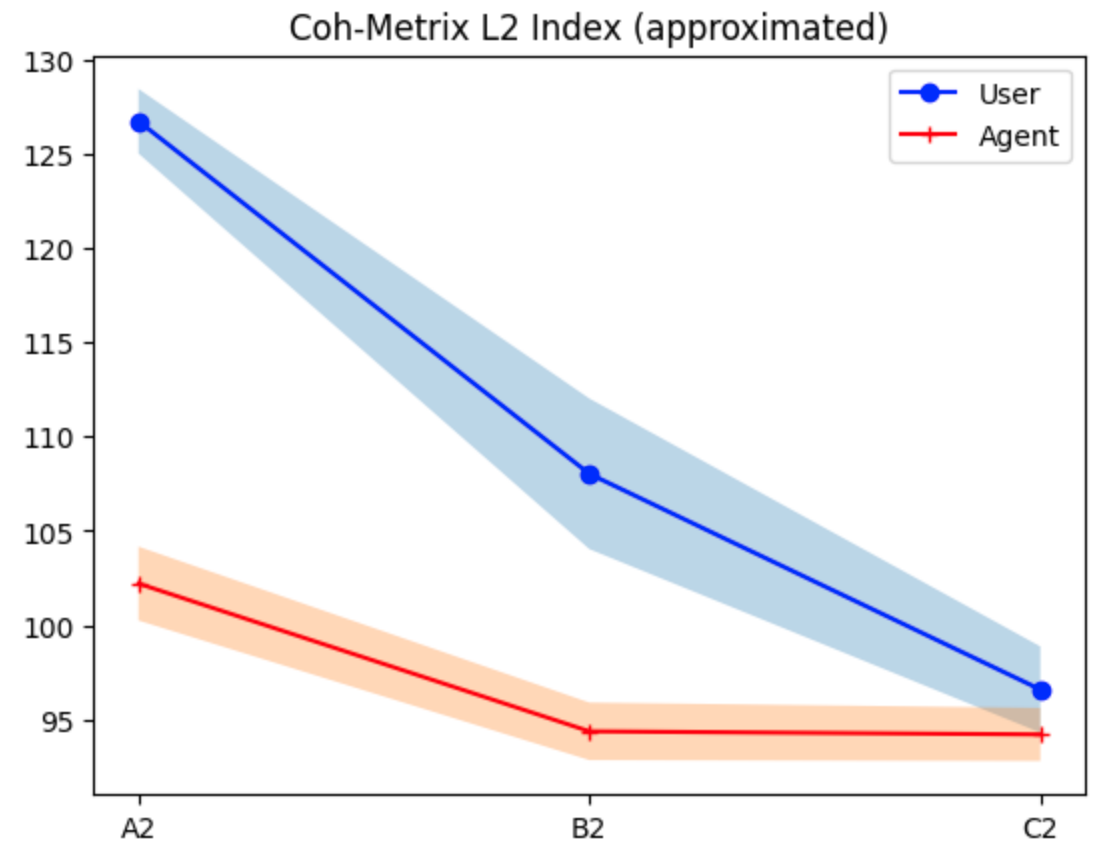}
    \caption{The CML2 average readability results for the A2, B2, and C2 CEFR levels  both for the User and the Agent}
    \label{fig:CML2}
\end{figure}

\begin{figure}
    \centering
    \includegraphics[scale=0.21]{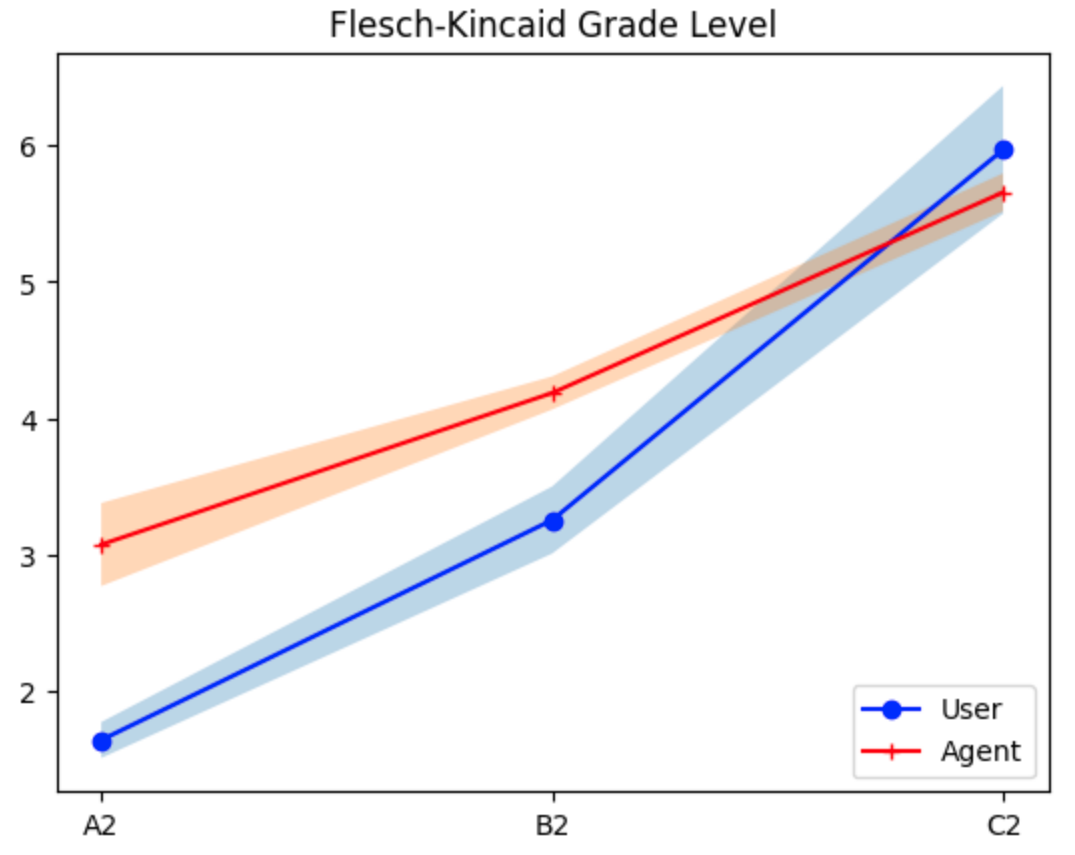}
    \caption{The FKG average readability results for the A2, B2, and C2 CEFR levels  both for the User and the Agent}
    \label{fig:FKG}
\end{figure}

\begin{figure}
    \centering
    \includegraphics[scale=0.21]{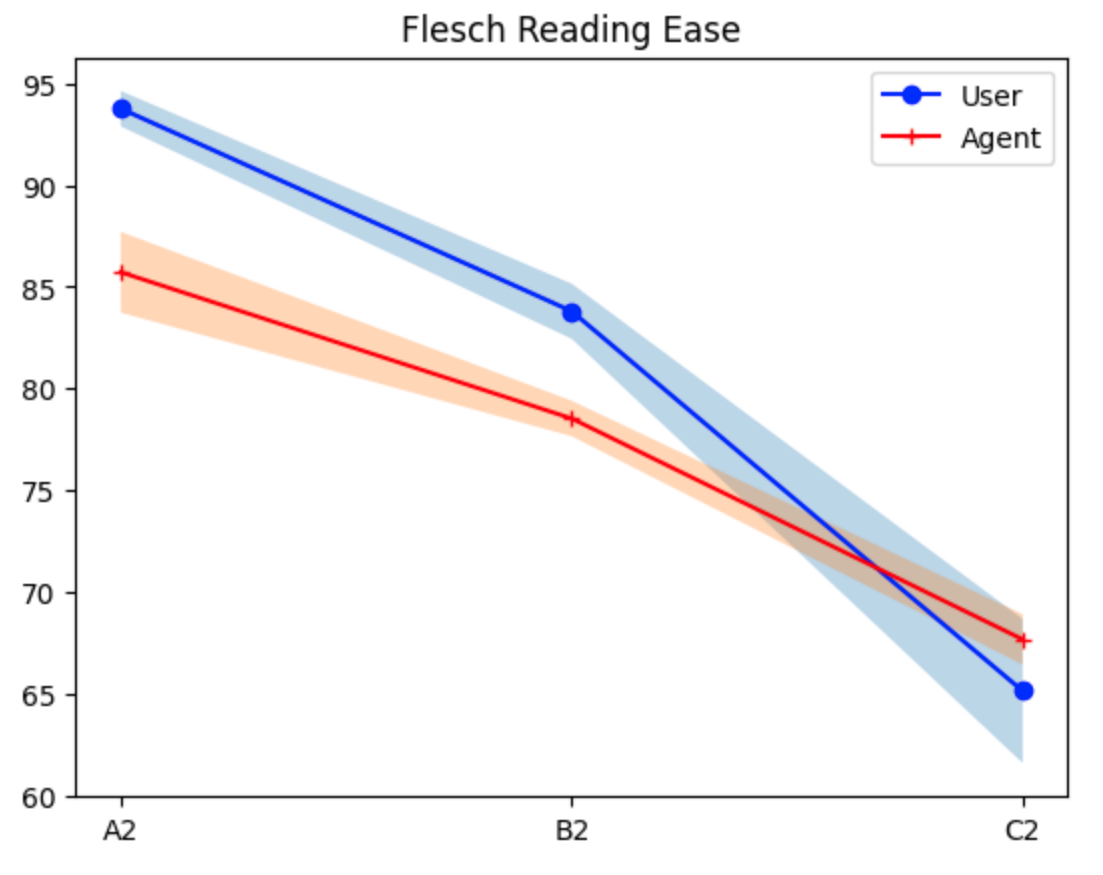}
    \caption{The FRE average readability results for the A2, B2, and C2 CEFR levels  both for the User and the Agent}
    \label{fig:FRE}
\end{figure}

\begin{figure}
    \centering
    \includegraphics[scale=0.21]{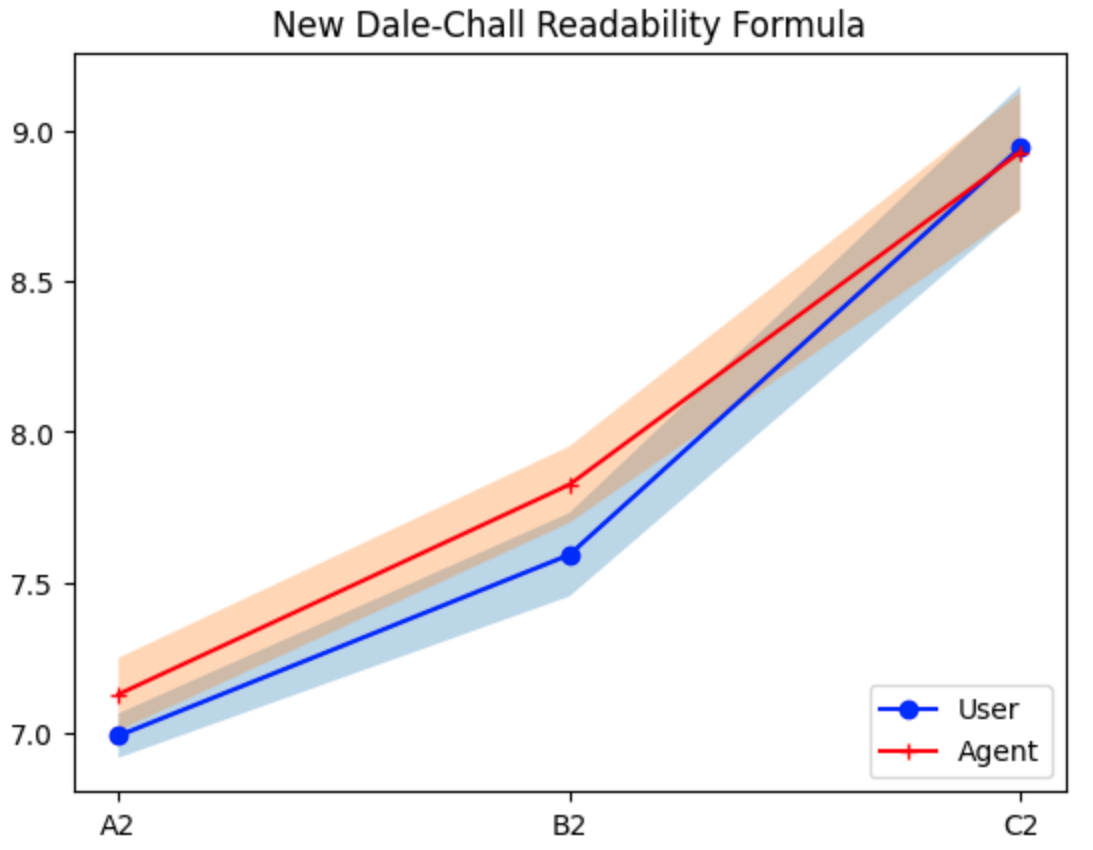}
    \caption{The NDC average readability results for the A2, B2, and C2 CEFR levels  both for the User and the Agent}
    \label{fig:NDC}
\end{figure}

\begin{figure}
    \centering
    \includegraphics[scale=0.21]{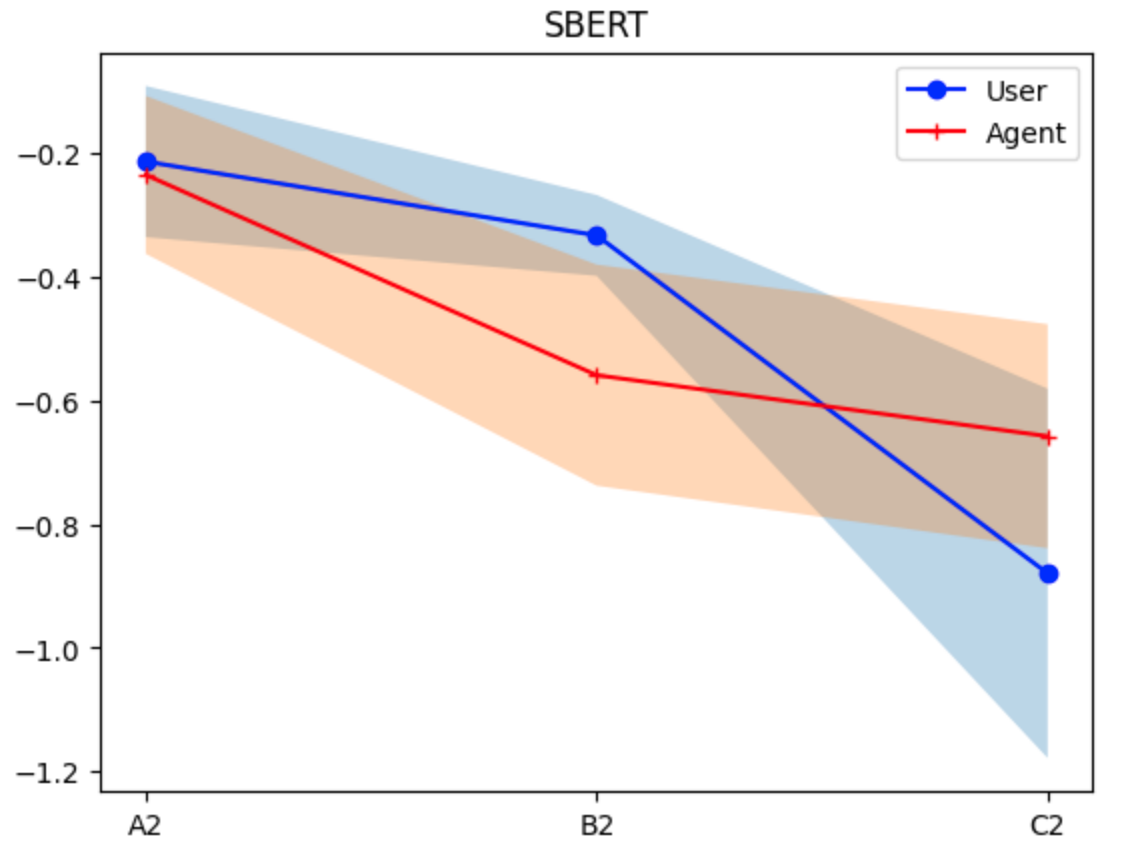}
    \caption{The SBERT average readability results for the A2, B2, and C2 CEFR levels both for the User and the Agent}
    \label{fig:SBERT}
\end{figure}

The readability results provided by the ARTE tool were then saved, and finally, an average value was computed. Moreover, we calculated the standard deviation of the measures taken. The results are illustrated in figures \ref{fig:ARI},  \ref{fig:CAREC}, \ref{fig:CARECM}, \ref{fig:CML2}, \ref{fig:FKG}, \ref{fig:FRE}, \ref{fig:NDC}, \ref{fig:SBERT}.

\begin{figure}
    \centering
    \includegraphics[scale=0.21]{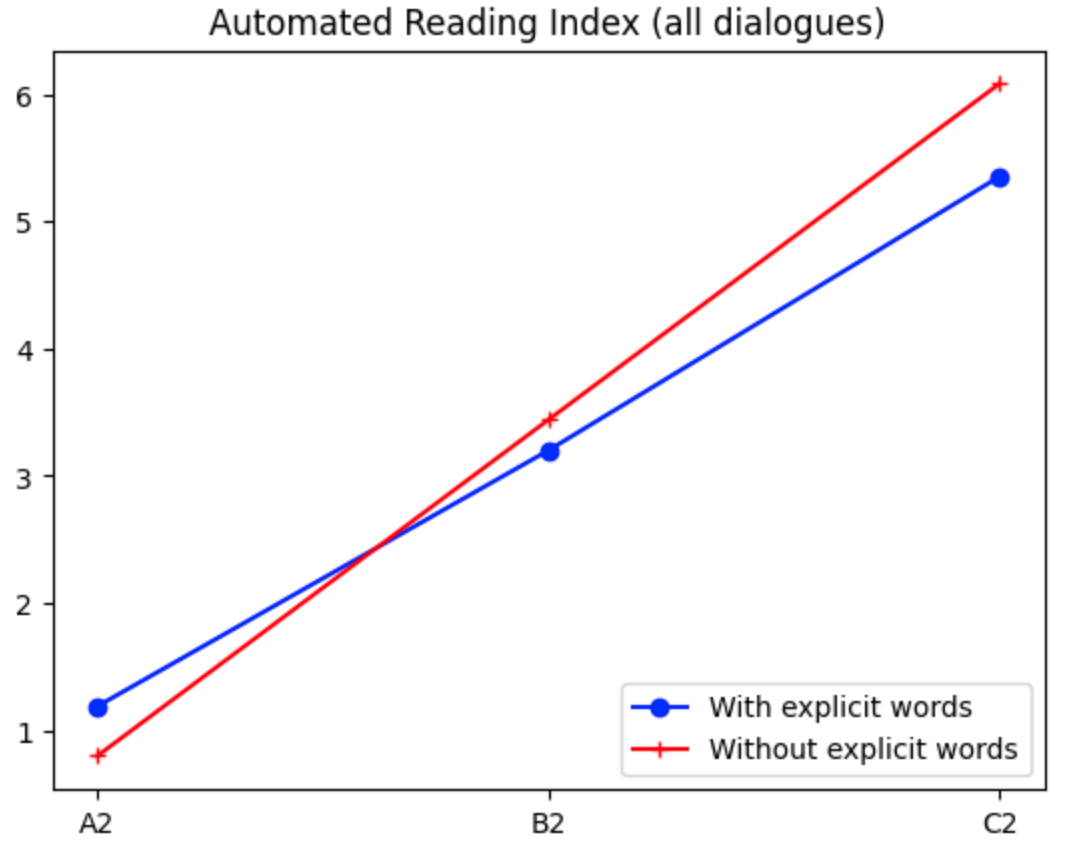}
    \caption{The ARI readability results for the A2, B2, and C2 CEFR levels considering the entire generated dialogues with or without words that explicitly refer to an emotion.}
    \label{fig:ARI-All}
\end{figure}

\begin{figure}
    \centering
    \includegraphics[scale=0.21]{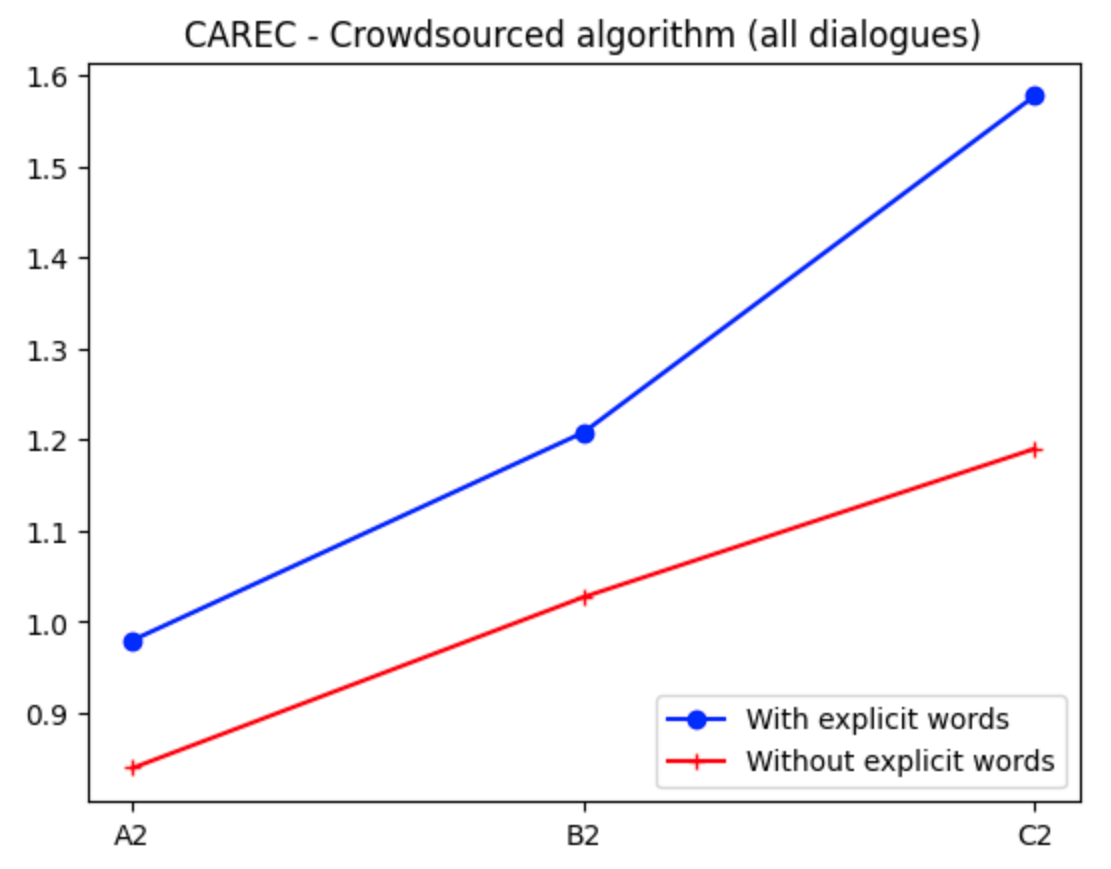}
    \caption{The CAREC readability results for the A2, B2, and C2 CEFR levels considering the entire generated dialogues with or without words that explicitly refer to an emotion.}
    \label{fig:CAREC-All}
\end{figure}

\begin{figure}
    \centering
    \includegraphics[scale=0.21]{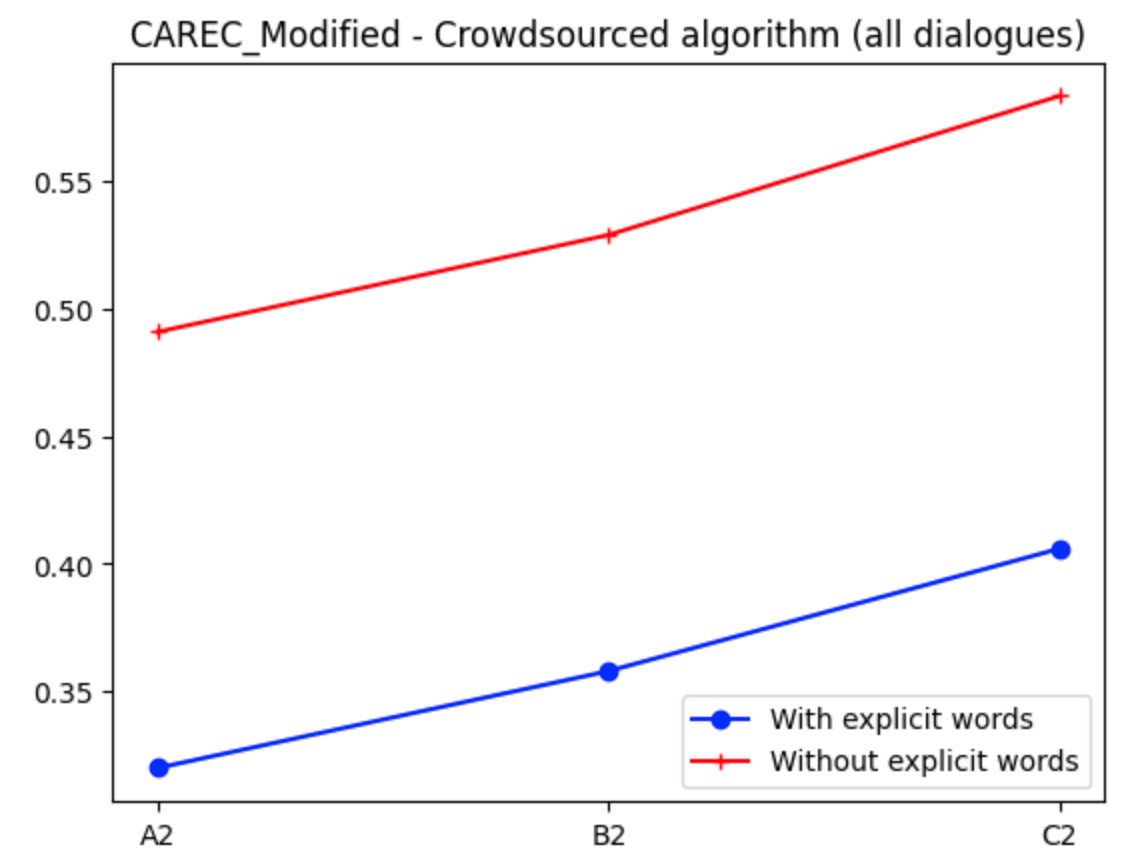}
    \caption{The CARECM readability results for the A2, B2, and C2 CEFR levels considering the entire generated dialogues with or without words that explicitly refer to an emotion.}
    \label{fig:CARECM-All}
\end{figure}

\begin{figure}
    \centering
    \includegraphics[scale=0.21]{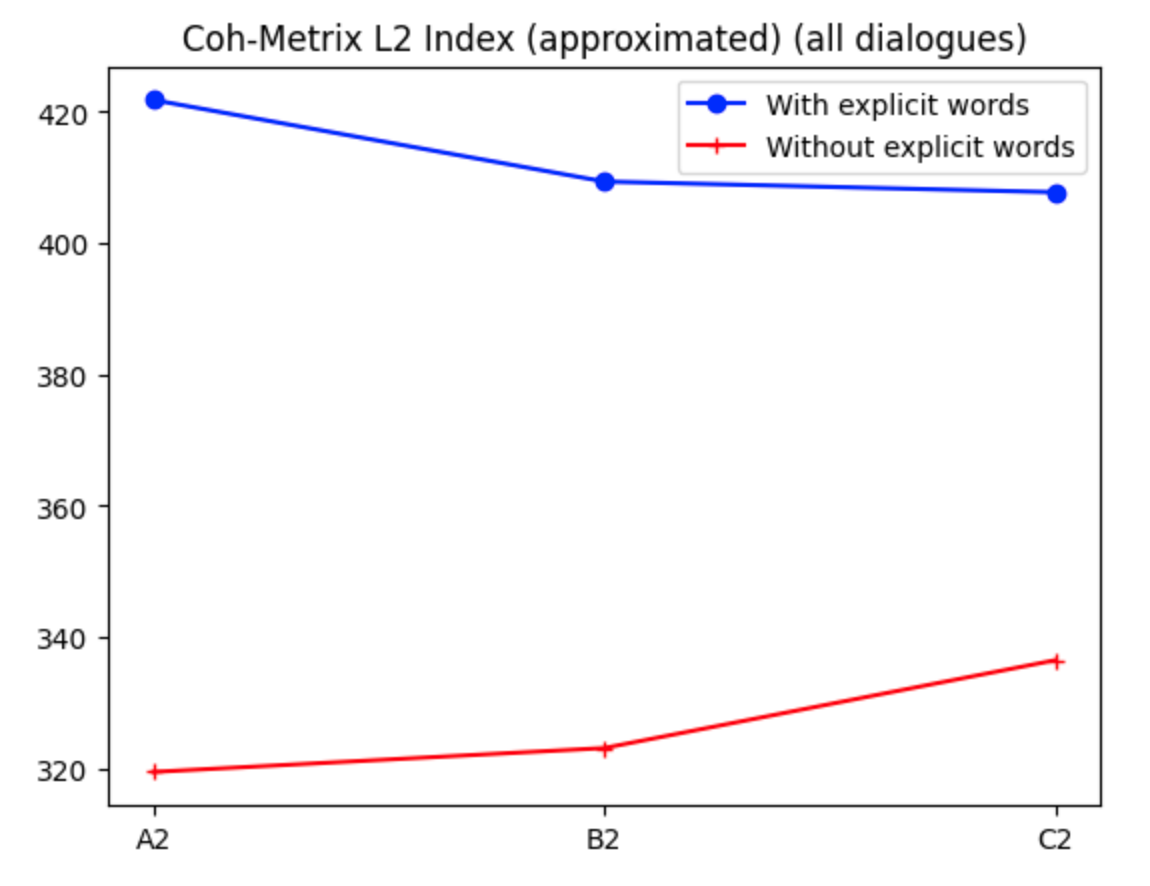}
    \caption{The CML2 readability results for the A2, B2, and C2 CEFR levels considering the entire generated dialogues with or without words that explicitly refer to an emotion.}
    \label{fig:CML2-All}
\end{figure}

\begin{figure}
    \centering
    \includegraphics[scale=0.21]{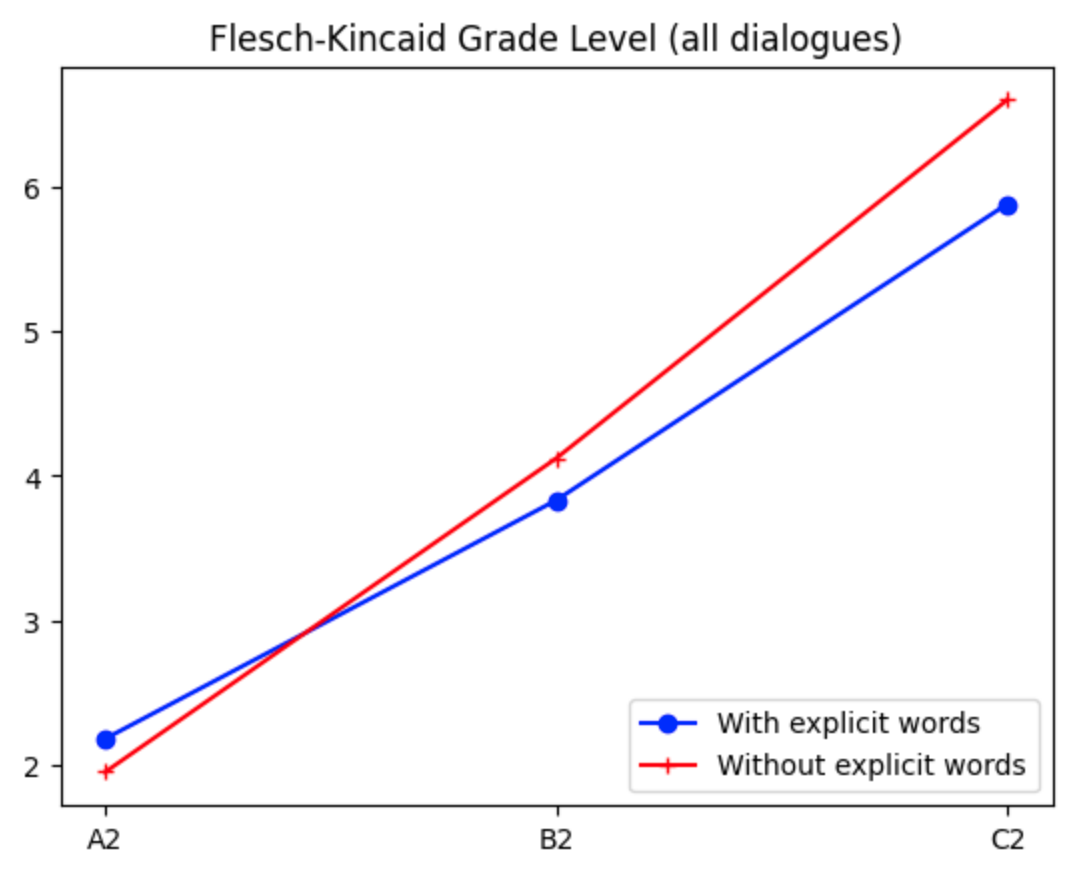}
    \caption{The FKG readability results for the A2, B2, and C2 CEFR levels considering the entire generated dialogues with or without words that explicitly refer to an emotion.}
    \label{fig:FKG-All}
\end{figure}

\begin{figure}
    \centering
    \includegraphics[scale=0.21]{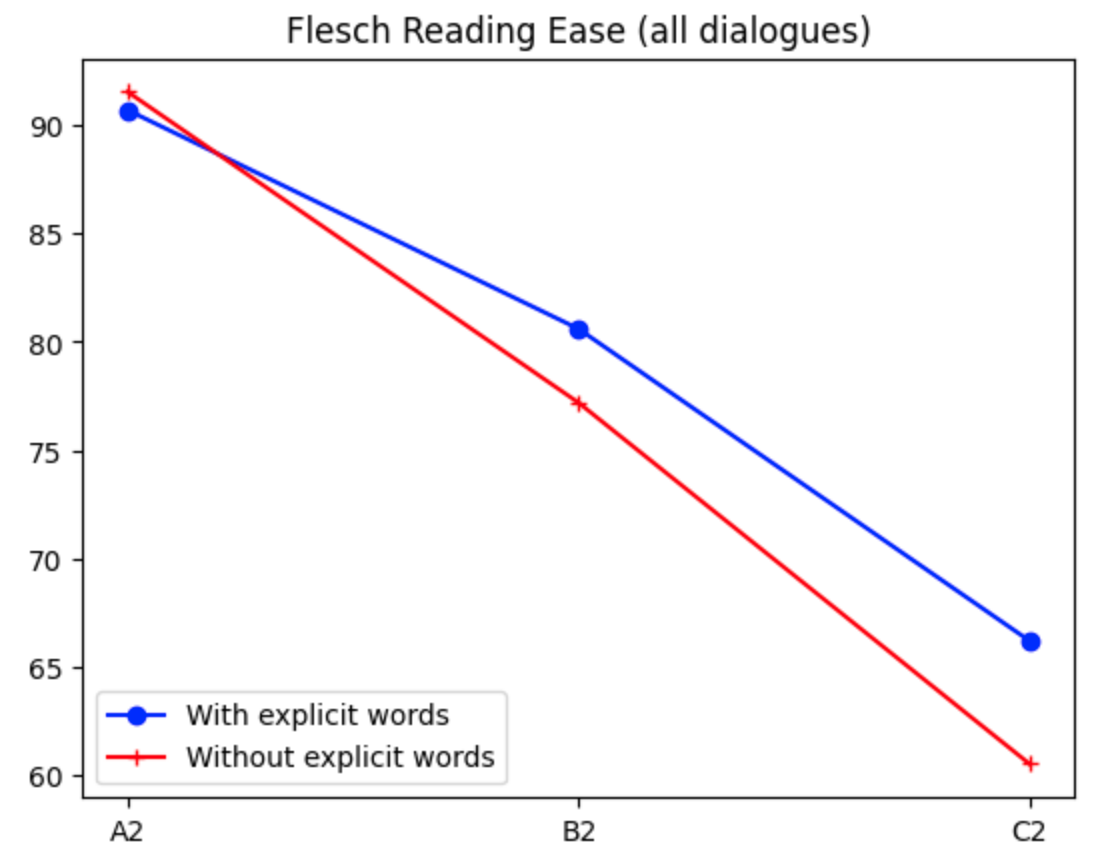}
    \caption{The FRE readability results for the A2, B2, and C2 CEFR levels considering the entire generated dialogues with or without words that explicitly refer to an emotion.}
    \label{fig:FRE-All}
\end{figure}

\begin{figure}
    \centering
    \includegraphics[scale=0.21]{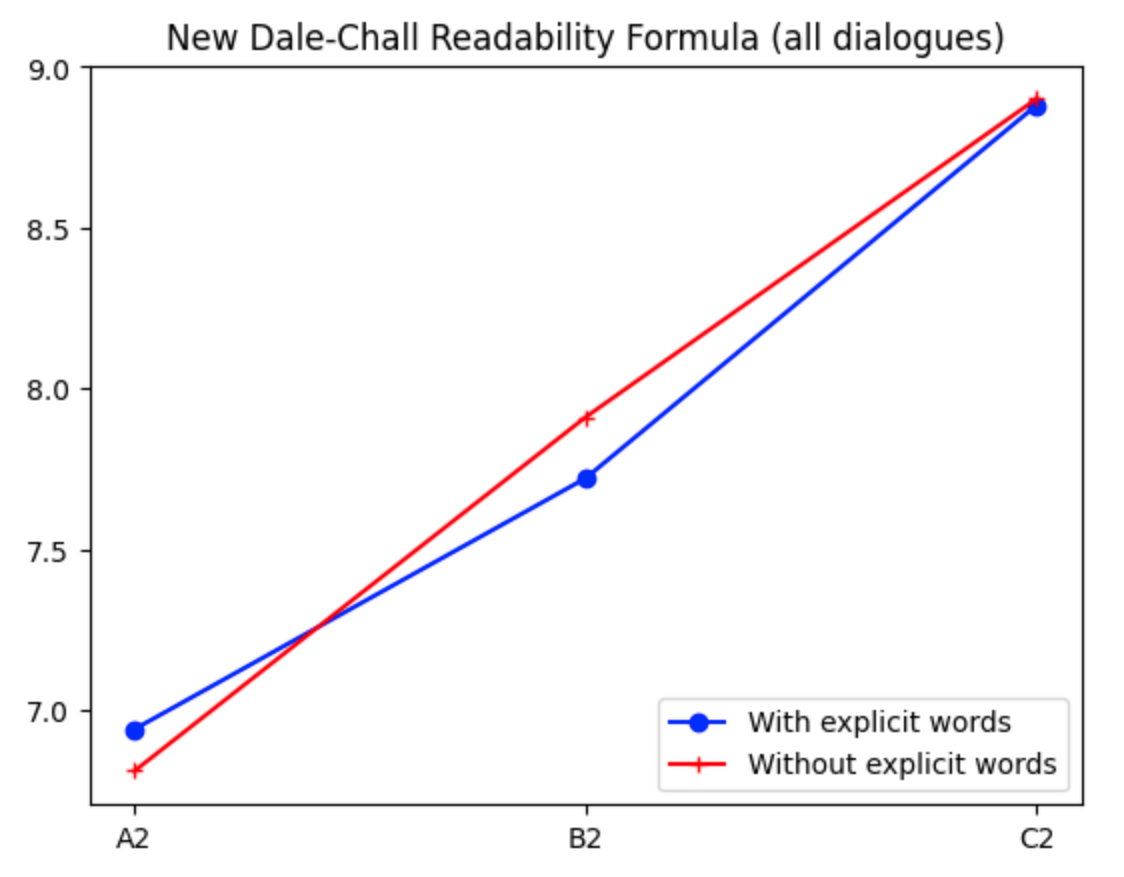}
    \caption{The NDC readability results for the A2, B2, and C2 CEFR levels considering the entire generated dialogues with or without words that explicitly refer to an emotion.}
    \label{fig:NDC-All}
\end{figure}

\begin{figure}
    \centering
    \includegraphics[scale=0.21]{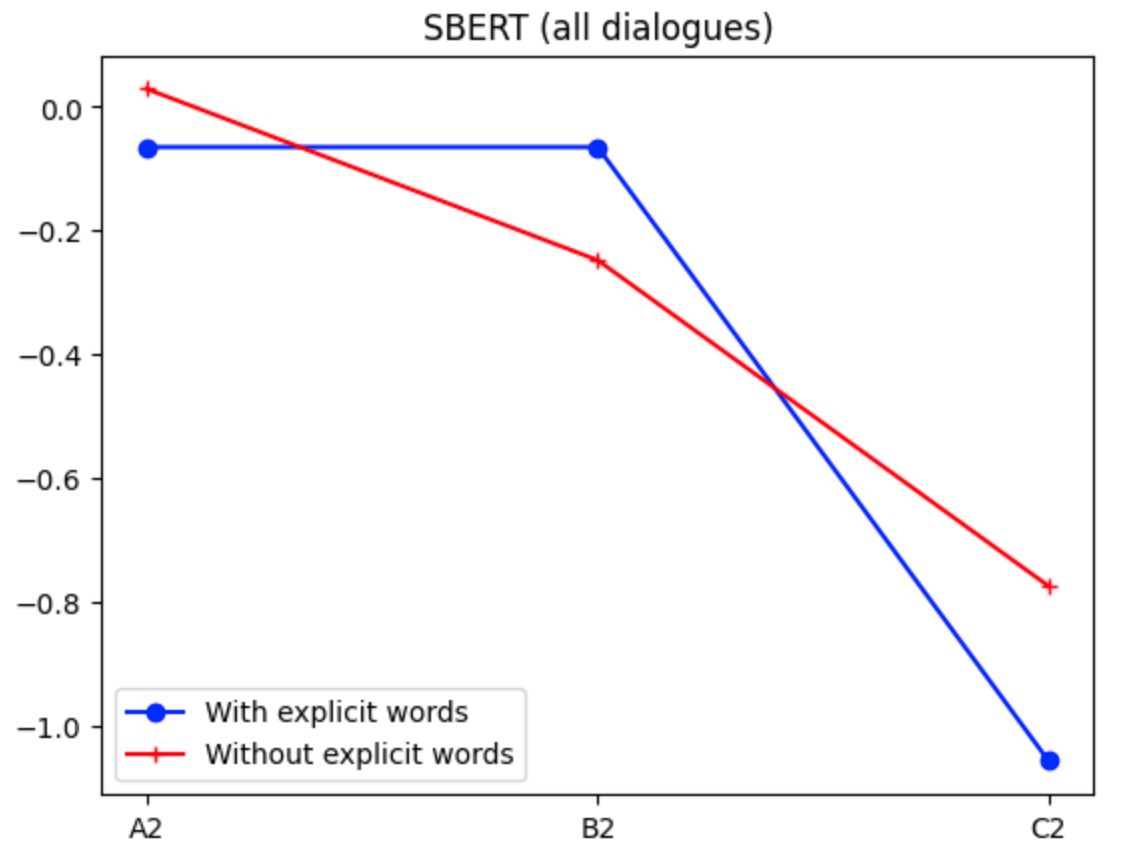}
    \caption{The SBERT readability results for the A2, B2, and C2 CEFR levels considering the entire generated dialogues with or without words that explicitly refer to an emotion.}
    \label{fig:SBERT-All}
\end{figure}

As a further analysis, we merged all the dialogues concerning a given CEFR level that explicitly used the emotion word; we did the same with those dialogues which did not explicitly used the emotion word. We executed readability tests on these types of files by using the ARTE tool \cite{ARTE}. The readability results for the \textit{A2}, \textit{B2}, and \textit{C2} CEFR levels considering the entire generated dialogues with or without words that explicitly refer to an emotion are reported in figures \ref{fig:ARI-All},  \ref{fig:CAREC-All}, \ref{fig:CARECM-All}, \ref{fig:CML2-All}, \ref{fig:FKG-All}, \ref{fig:FRE-All}, \ref{fig:NDC-All}, \ref{fig:SBERT-All}.

\begin{figure}
    \centering
    \includegraphics[scale=0.21]{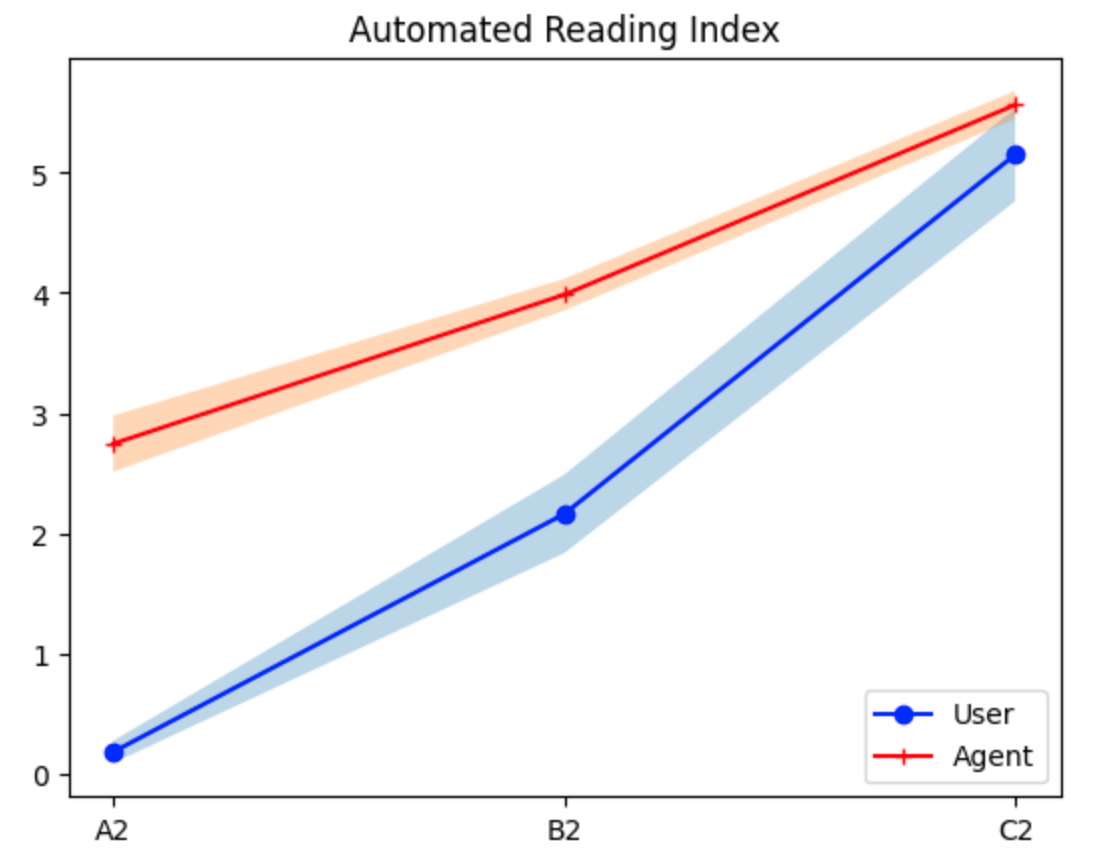}
    \caption{The ARI average readability results for the A2, B2, and C2 CEFR levels  both for the User and the Agent}
    \label{fig:ARI}
\end{figure}

\section{Conclusions and Future work}

We have developed an analytics tool to construct a dataset of dialogues between users and agents within a customer care setting. This system leverages ChatGPT 3.5 to generate hypothetical dialogues within a specified context, considering two critical factors: the emotional state expressed by the user and the linguistic complexity employed during the interaction.

The dialogues have also been evaluated by using different parameters to verify that they met an acceptable quality level. Our tool can be helpful, for example, in reinforcement learning applications, where a satisfaction score could be provided after each conversation, and in facilitating effective human-computer interactions and conversation management within specific contexts.

The results of our study have highlighted both advantages and limitations associated with the use of automatic dialogue generation systems. The primary benefit lies in the accelerated rate of data generation. Another significant advantage is the ability to create novel conversations from scratch, prompting ChatGPT to generate responses that adhere to pre-defined criteria.

Of course it is important to verify the quality of generated dialogues. The outcomes of our readability measurements have confirmed the efficacy of ChatGPT in producing dialogues characterized by diverse language levels and the possibility of leveraging attitude tags associated with each turn of conversation to infer effective interaction habits. 

Future research will focus on three key areas: \textit{(1)} enhancing our analytics system to integrate it into decision support systems for human-computer and human-robot interaction (e.g., \cite{DBLP:journals/eaai/Maroto-GomezMCG23,DBLP:journals/ijhr/HanHEH23}); \textit{(2)} exploring the intersection of our research with emerging trends in big data analysis (e.g., \cite{DBLP:conf/dawak/CuzzocreaRS08,DBLP:journals/dke/CuzzocreaC10,DBLP:journals/jiis/CeciCM15}); \textit{(3)} focusing on interesting adaptive and user-personalization metaphors, perhaps developed in related scientific areas (e.g., \cite{DBLP:conf/icdt/0001C001,DBLP:conf/www/0001CMO002,DBLP:journals/wias/Cuzzocrea06}).  

\section*{Acknowledgements} This work was funded by the Next Generation EU - Italian NRRP, Mission 4, Component 2, Investment 1.5 (Directorial Decree n. 2021/3277) - project Tech4You n. ECS0000009.

\bibliographystyle{IEEEtran}

\begin{thebibliography}{10}
\providecommand{\url}[1]{#1}
\csname url@samestyle\endcsname
\providecommand{\newblock}{\relax}
\providecommand{\bibinfo}[2]{#2}
\providecommand{\BIBentrySTDinterwordspacing}{\spaceskip=0pt\relax}
\providecommand{\BIBentryALTinterwordstretchfactor}{4}
\providecommand{\BIBentryALTinterwordspacing}{\spaceskip=\fontdimen2\font plus
\BIBentryALTinterwordstretchfactor\fontdimen3\font minus \fontdimen4\font\relax}
\providecommand{\BIBforeignlanguage}[2]{{%
\expandafter\ifx\csname l@#1\endcsname\relax
\typeout{** WARNING: IEEEtran.bst: No hyphenation pattern has been}%
\typeout{** loaded for the language `#1'. Using the pattern for}%
\typeout{** the default language instead.}%
\else
\language=\csname l@#1\endcsname
\fi
#2}}
\providecommand{\BIBdecl}{\relax}
\BIBdecl

\bibitem{pilato2018data}
G.~Pilato and E.~D’Avanzo, ``Data-driven social mood analysis through the conceptualization of emotional fingerprints,'' \emph{Procedia computer science}, vol. 123, pp. 360--365, 2018.

\bibitem{cuzzocrea2021composite}
A.~Cuzzocrea and G.~Pilato, ``A composite framework for supporting user emotion detection based on intelligent taxonomy handling,'' \emph{Logic Journal of the IGPL}, vol.~29, no.~2, pp. 207--219, 2021.

\bibitem{moin2023emotion}
A.~Moin, F.~Aadil, Z.~Ali, and D.~Kang, ``Emotion recognition framework using multiple modalities for an effective human--computer interaction,'' \emph{The Journal of Supercomputing}, vol.~79, no.~8, pp. 9320--9349, 2023.

\bibitem{poria2019emotion}
S.~Poria, N.~Majumder, R.~Mihalcea, and E.~Hovy, ``Emotion recognition in conversation: Research challenges, datasets, and recent advances,'' \emph{IEEE Access}, vol.~7, pp. 100\,943--100\,953, 2019.

\bibitem{ekman1994all}
P.~Ekman, ``All emotions are basic,'' \emph{The nature of emotion: Fundamental questions}, pp. 15--19, 1994.

\bibitem{Crowdflower2016}
Crowdflower, ``The emotion in text. https://www.figure-eight.com/data/ sentiment-analysis-emotion-text/.'' 2016.

\bibitem{joshi-etal-2016-harnessing}
A.~Joshi, V.~Tripathi, P.~Bhattacharyya, and M.~J. Carman, ``Harnessing sequence labeling for sarcasm detection in dialogue from {TV} series {`}{F}riends{'},'' in \emph{Proceedings of the 20th {SIGNLL} {CoNLL} Conference}, 2016, pp. 146--155.

\bibitem{buechel-hahn-2017-emobank}
S.~Buechel and U.~Hahn, ``{E}mo{B}ank: Studying the impact of annotation perspective and representation format on dimensional emotion analysis,'' in \emph{Proceedings of the 15th {EACL} Conference}, 2017, pp. 578--585.

\bibitem{SemEval2018Task1}
S.~Mohammad, F.~Bravo-Marquez, M.~Salameh, and S.~Kiritchenko, ``{S}em{E}val-2018 task 1: Affect in tweets,'' in \emph{Proceedings of the 12th International Workshop on Semantic Evaluation}, 2018, pp. 1--17.

\bibitem{goel2016real}
A.~Goel, J.~Gautam, and S.~Kumar, ``Real time sentiment analysis of tweets using naive bayes,'' in \emph{2016 2nd {NGCT} International Conference}.\hskip 1em plus 0.5em minus 0.4em\relax IEEE, 2016, pp. 257--261.

\bibitem{mohammad-bravo-marquez-2017-wassa}
S.~Mohammad and F.~Bravo-Marquez, ``{WASSA}-2017 shared task on emotion intensity,'' in \emph{Proceedings of the 8th WASSA Workshop}, 2017, pp. 34--49.

\bibitem{scherer1994evidence}
K.~R. Scherer and H.~G. Wallbott, ``Evidence for universality and cultural variation of differential emotion response patterning: Correction.'' \emph{J. Pers. Soc. Psychology}, 1994.

\bibitem{schuff2017annotation}
H.~Schuff, J.~Barnes, J.~Mohme, S.~Pad{\'o}, and R.~Klinger, ``Annotation, modelling and analysis of fine-grained emotions on a stance and sentiment detection corpus,'' in \emph{Proceedings of the 8th WASSA Workshop}, 2017, pp. 13--23.

\bibitem{mohammad2017stance}
S.~M. Mohammad, P.~Sobhani, and S.~Kiritchenko, ``Stance and sentiment in tweets,'' \emph{ACM Transactions on Internet Technology (TOIT)}, vol.~17, no.~3, pp. 1--23, 2017.

\bibitem{koelstra2011deap}
S.~Koelstra, C.~Muhl, M.~Soleymani, J.-S. Lee, A.~Yazdani, T.~Ebrahimi, T.~Pun, A.~Nijholt, and I.~Patras, ``Deap: A database for emotion analysis; using physiological signals,'' \emph{IEEE transactions on affective computing}, vol.~3, no.~1, pp. 18--31, 2011.

\bibitem{busso2008iemocap}
C.~Busso, M.~Bulut, C.-C. Lee, A.~Kazemzadeh, E.~Mower, S.~Kim, J.~N. Chang, S.~Lee, and S.~S. Narayanan, ``Iemocap: Interactive emotional dyadic motion capture database,'' \emph{Language resources and evaluation}, vol.~42, pp. 335--359, 2008.

\bibitem{li2017dailydialog}
Y.~Li, H.~Su, X.~Shen, W.~Li, Z.~Cao, and S.~Niu, ``Dailydialog: A manually labelled multi-turn dialogue dataset,'' in \emph{Proceedings of the 8th International Joint Conference on Natural Language Processing}, 2017, pp. 986--995.

\bibitem{li-etal-2016-deep}
J.~Li, W.~Monroe, A.~Ritter, D.~Jurafsky, M.~Galley, and J.~Gao, ``Deep reinforcement learning for dialogue generation,'' in \emph{Proceedings of the 2016 Conference on Empirical Methods in Natural Language Processing}, J.~Su, K.~Duh, and X.~Carreras, Eds., 2016, pp. 1192--1202.

\bibitem{steindl2023generating}
S.~Steindl, U.~Sch{\"a}fer, and B.~Ludwig, ``Generating synthetic dialogues from prompts to improve task-oriented dialogue systems,'' in \emph{German Conference on Artificial Intelligence}.\hskip 1em plus 0.5em minus 0.4em\relax Springer, 2023, pp. 207--214.

\bibitem{chen2023places}
M.~Chen, A.~Papangelis, C.~Tao, S.~Kim, A.~Rosenbaum, Y.~Liu, Z.~Yu, and D.~Hakkani-Tur, ``Places: Prompting language models for social conversation synthesis,'' in \emph{Findings of the Association for Computational Linguistics: EACL 2023}, 2023, pp. 844--868.

\bibitem{li2022controllable}
Z.~Li, W.~Chen, S.~Li, H.~Wang, J.~Qian, and X.~Yan, ``Controllable dialogue simulation with in-context learning,'' in \emph{Findings of the Association for Computational Linguistics: EMNLP 2022}, 2022, pp. 4330--4347.

\bibitem{mehri2022lad}
S.~Mehri, Y.~Altun, and M.~Eskenazi, ``Lad: Language models as data for zero-shot dialog,'' in \emph{Proceedings of the 23rd Annual Meeting of the Special Interest Group on Discourse and Dialogue}, 2022, pp. 595--604.

\bibitem{kulkarni2024synthdst}
A.~Kulkarni, B.-H. Tseng, J.~Moniz, D.~Piraviperumal, H.~Yu, and S.~Bhargava, ``Synthdst: Synthetic data is all you need for few-shot dialog state tracking,'' in \emph{Proceedings of the 18th {EACL} Conference}, 2024, pp. 1988--2001.

\bibitem{ekman1999basic}
P.~Ekman \emph{et~al.}, ``Basic emotions,'' \emph{Handbook of cognition and emotion}, vol.~98, no. 45-60, p.~16, 1999.

\bibitem{asta2023supporting}
A.~Asta, A.~Cuzzocrea, A.~Fantini, G.~Pilato, and P.~G. Bringas, ``Supporting emotion recognition in human-robot interactions: An experimental italian textual dataset,'' in \emph{International Conference on Hybrid Artificial Intelligence Systems}.\hskip 1em plus 0.5em minus 0.4em\relax Springer, 2023, pp. 483--494.

\bibitem{ARTE}
ARTE, ``Automatic readability tool for english, https://www.linguisticanalysistools.org/arte.html.''

\bibitem{kincaid1975derivation}
J.~P. Kincaid, R.~P. Fishburne~Jr, R.~L. Rogers, and B.~S. Chissom, ``Derivation of new readability formulas (automated readability index, fog count and flesch reading ease formula) for navy enlisted personnel,'' \emph{IST Technical Report 8-75}, 1975.

\bibitem{chall1995readability}
J.~S. Chall and E.~Dale, ``Readability revisited: The new dale-chall readability formula,'' \emph{(No Title)}, 1995.

\bibitem{reimers2019sentence}
N.~Reimers and I.~Gurevych, ``Sentence-bert: Sentence embeddings using siamese bert-networks,'' in \emph{Proceedings of the 2019 EMNLP-IJCNLP Joint Conferences}, 2019, pp. 3982--3992.

\bibitem{DBLP:journals/eaai/Maroto-GomezMCG23}
M.~Maroto{-}G{\'{o}}mez, S.~Marques{-}Villarroya, J.~C. Castillo, {\'{A}}.~C. Gonz{\'{a}}lez, and M.~Malfaz, ``Active learning based on computer vision and human-robot interaction for the user profiling and behavior personalization of an autonomous social robot,'' \emph{Eng. Appl. Artif. Intell.}, vol. 117, no. Part, p. 105631, 2023.

\bibitem{DBLP:journals/ijhr/HanHEH23}
X.~Han, D.~Huang, S.~Eun{-}Lee, and J.~Hoon{-}Yang, ``Artificial intelligence-oriented user interface design and human behavior recognition based on human-computer nature interaction,'' \emph{Int. J. Humanoid Robotics}, vol.~20, no.~6, pp. 2\,250\,020:1--2\,250\,020:19, 2023.

\bibitem{DBLP:conf/dawak/CuzzocreaRS08}
A.~Cuzzocrea, V.~Russo, and D.~Sacc{\`{a}}, ``A robust sampling-based framework for privacy preserving {OLAP},'' in \emph{Proceedings of 10th DaWaK International Conference}.\hskip 1em plus 0.5em minus 0.4em\relax Springer, 2008, pp. 97--114.

\bibitem{DBLP:journals/dke/CuzzocreaC10}
A.~Cuzzocrea and S.~Chakravarthy, ``Event-based lossy compression for effective and efficient {OLAP} over data streams,'' \emph{Data Knowl. Eng.}, vol.~69, no.~7, pp. 678--708, 2010.

\bibitem{DBLP:journals/jiis/CeciCM15}
M.~Ceci, A.~Cuzzocrea, and D.~Malerba, ``Effectively and efficiently supporting roll-up and drill-down {OLAP} operations over continuous dimensions via hierarchical clustering,'' \emph{J. Intell. Inf. Syst.}, vol.~44, no.~3, pp. 309--333, 2015.

\bibitem{DBLP:conf/icdt/0001C001}
M.~Cannataro, A.~Cuzzocrea, and A.~Pugliese, ``A probabilistic approach to model adaptive hypermedia systems,'' in \emph{Proceedings of the First WebDyn International Workshop @ICDT2001}, 2001, pp. 50--60.

\bibitem{DBLP:conf/www/0001CMO002}
M.~Cannataro, A.~Cuzzocrea, C.~Mastroianni, R.~Ortale, and A.~Pugliese, ``Modeling adaptive hypermedia with an object-oriented approach and {XML},'' in \emph{Proceedings of the Second WebDyn International Workshop @ WWW2002}, 2002, pp. 35--44.

\bibitem{DBLP:journals/wias/Cuzzocrea06}
A.~Cuzzocrea, ``Combining multidimensional user models and knowledge representation and management techniques for making web services knowledge-aware,'' \emph{Web Intell. Agent Syst.}, vol.~4, no.~3, pp. 289--312, 2006.

\end{thebibliography}


\end{document}